\documentclass[conference]{IEEEtran}
\IEEEoverridecommandlockouts
\usepackage{cite}
\usepackage{amsmath,amssymb,amsfonts}
\usepackage{algorithmic}
\usepackage{graphicx}
\usepackage{textcomp}
\usepackage{xcolor}
\usepackage{subfigure}
\usepackage{verbatim}

\def\BibTeX{{\rm B\kern-.05em{\sc i\kern-.025em b}\kern-.08em
    T\kern-.1667em\lower.7ex\hbox{E}\kern-.125emX}}
\begin{document}

\title{Information-Theoretic Limits on Compression of Semantic Information \\}

\author{\IEEEauthorblockN{Jiancheng Tang, Qianqian Yang\IEEEauthorrefmark{2}, Zhaoyang Zhang}
\IEEEauthorblockA{College of information Science and Electronic Engineering, Zhejiang University, Hangzhou 310007, China\\
Email: \{jianchengtang,qianqianyang20\IEEEauthorrefmark{2},ning\_ming\}@zju.edu.cn
}
\thanks{{\thefootnote}{*}This work is partly supported by the SUTD-ZJU IDEA Grant (SUTD-ZJU (VP) 202102), and partly by the Fundamental Research Funds for the Central Universities under Grant 2021FZZX001-20.}
}

\maketitle

\begin{abstract}

As conventional communication systems based on classic information theory have closely approached the limits of Shannon channel capacity, semantic communication has been recognized as a key enabling technology for the further improvement of communication performance. However, it is still unsettled on how to represent semantic information and characterise the theoretical limits. In this paper, we consider a semantic source which consists of a set of correlated random variables whose joint probabilistic distribution can be described by a Bayesian network. Then we give the information-theoretic limit on the lossless compression of the semantic source and introduce a low complexity encoding method by exploiting the conditional independence. We further characterise the limits on lossy compression of the semantic source and the corresponding upper and lower bounds of the rate-distortion function. We also investigate the lossy compression of the semantic source with side information at both the encoder and decoder, and obtain the rate distortion function. We prove that the optimal code of the semantic source is the combination of the optimal codes of each conditional independent set given the side information.

 
\end{abstract}

\section{Introduction}
The classical information theory (CIT) established by Shannon in 1948 is the cornerstone of modern communication systems. Concentrating on the accurate symbol transmission while ignoring the semantic content of communications, Shannon defined the information entropy based on the probabilistic distribution of symbols to measure the size of information quantitatively \cite{b1}, based on which the theoretical limits on source compression and channel capacity are characterised. With the development of digital communications over the past 70 years, existing communication techniques, such as polar code and multiple-input multiple-output (MIMO) systems, have pushed the current communication systems closely approaching the Shannon capacity\cite{b2}\cite{b3}. To further improve the communication efficiency in order to meet the ever-growing demands, semantic oriented communication has attracted a lot of research interest lately, and widely recognized as a promising approach to overcome the Shannon limits \cite{b4,overview1,overview2,overview3}. 

Different from the traditional communication approaches, semantic communication systems only transmit the semantic or task relevant information while remove the redundancy to improve transmission efficiency\cite{b8,overview4,overview5,overview6}. Semantic oriented communication methods have been implemented based on deep learning techniques for the efficient transmission of image \cite{b9.1,image1,image2,image3,image4}, text \cite{text1,text2}, video \cite{video1,video2} and speech signals \cite{b9,speech2,speech3}. These methods have been shown to achieve higher transmission efficiency compared with conventional methods for the specific tasks they are designed for. Despite this success, the design of semantic communication system still lacks theoretical guidance. 

The research on semantic information theory can date back to about the time when the classical information theory was proposed. In one of a few early works\cite{b14,b16}, Carnap and Bar Hillel proposed to use propositional logic sentences to represent semantic information. The semantic information entropy is calculated based on logical probabilities \cite{b15}, instead of statistical probability as in classical information theory.  Bao \emph{et al.}\cite{Bao} further extended this theoretical work and derived the semantic channel capacity of discrete memoryless channel based on propositional logic probabilities. De Luca \emph{et al.} \cite{b12,b13} denoted semantic information by fuzzy variable and introduced fuzzy entropy to measure the uncertainty of fuzzy variables. However, neither the propositional logic nor fuzzy variables are expressive enough to describe semantic information of the complex data in today's applications. 


Recently, Liu \emph{et al.} proposed a new source model, where they viewed its semantic information as an intrinsic part of the source that is not observable but can be inferred from the extrinsic state\cite{Poor}. They characterised the defined the semantic rate-distortion function through classical indirect  rate-distortion theory based on this source model. Similarly, Guo \emph{et al.} also modeled the semantic information as the unobservable information in a source, and characterized the theoretic limits on the rate distortion problem with side information \cite{Guo}. In  \cite{Shao},  the authors argued that the design of semantic language that maps meaning to messages is essentially a joint source-channel coding problem and characterised the trade-off between the rate and a general distortion measure. These works have shed light on developing a generic theory of semantic communication. However, the inner structure of semantic information remains unexplored. 

In this paper, we consider a semantic source as a set of correlated semantic elements whose joint distribution can be modeled by a Bayesian network (BN). We characterise the information-theoretic limits on the lossless compression and lossy compression of semantic sources and derive the lower and upper bounds on the rate-distortion function. We further study the lossy compression problem with side information at both sides and prove that the optimality of compressing each conditionally independent set of variables given the side information. We derive the conditional rate-distortion functions when semantic elements follow binary or Gaussian distributed.

The organization of the rest of the paper is as follows: we introduce the semantic source in Section II. In Section III, we discuss information-theoretic limits on the compression of semantic source. In section IV, we study the problem of lossy compression with two-sided state information. In Section V, we conclude the paper.

\section{Semantic source model and Semantic Communication system}
\begin{figure}{}
\centering
\subfigure[]{\includegraphics[width=3cm]{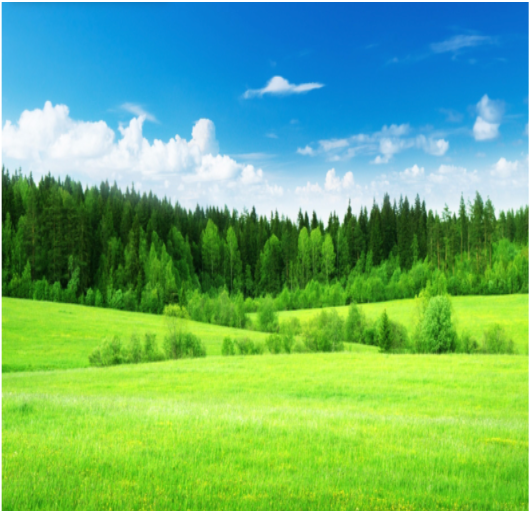}} 
\subfigure[]{\includegraphics[width=4cm]{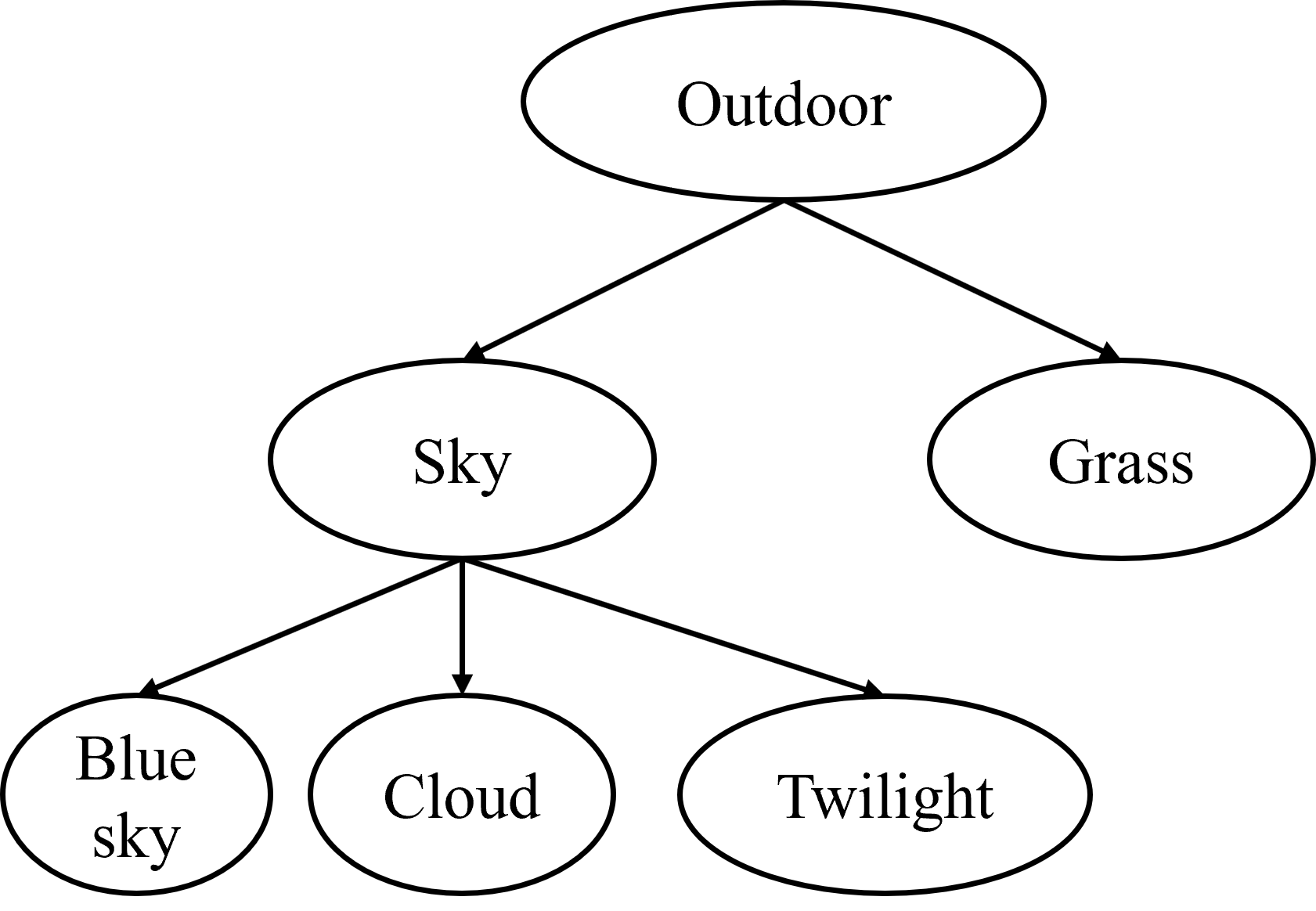}}
\caption{(a) The original image. (b) The BN model of semantic elements of the image.} 
\label{fig2}
\end{figure}

\begin{figure}
    \centering
	\includegraphics[width=3in]{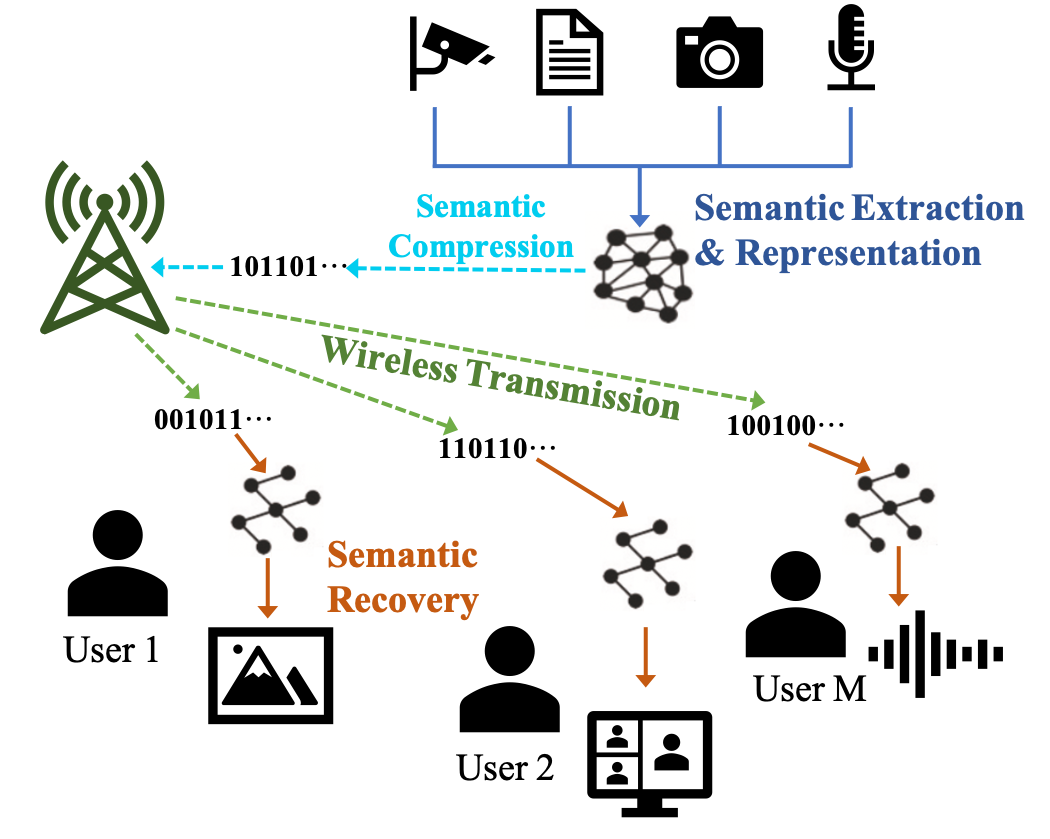}
	\caption{ The BN-enabled semantic communication framework. }
	\label{structure}
\end{figure}
In this paper, we assume that a semantic source consists of a set of correlated semantic elements whose joint probabilistic distribution is modeled by a BN. BN has been widely used in semantic analysis and understanding of various types of data \cite{b21,b22,b222}. For example, Luo \emph{et al}. proposed a scene classification method of images in which the semantic features are represented by a set of correlated semantic elements \cite{b24}. 
An image and the BN model of its semantic elements are shown in Fig.~\ref{fig2}(a) and Fig.~\ref{fig2}(b), where each node in  Fig.~\ref{fig2}(b) represents a semantic element. The conditional dependence relations among the semantic elements are obtained by expert knowledge, and the conditional probability matrices (CPMs) of each node are obtained by using the frequency counting approach based on an image dataset. For example, the semantic features \emph{sky} and \emph{grass} are extracted by an object detection algorithm and used as evidences to determine the scene category. In particular, the image is detected as \emph{outdoor} when the posterior probability of the root node is large than a predefined threshold. In addition to the image procession, BN has also widely applied to various tasks representing semantic relations in different type of data such as text \cite{b23} and videos \cite{b25}.
The BN-enabled semantic communication framework is shown in Fig.~\ref{structure}, which consists of  four phases: a) semantic extraction and representation, b) semantic compression, c) semantic transmission, and d) original data recovery. In this paper, we assume that the conventional information source has been converted into the semantic source by using modern deep learning techniques in semantic extraction and representation phase. Our focus is on the semantic compression phase, where we provide some information-theoretic limits for compressing the semantic source.


\section{Semantic Compression For Correlated Semantic Elements}
In this section, we present the theoretical limits on lossless and lossy  compression of semantic sources, i.e., a set of correlated semantic elements whose correlation are modeled by BNs. We consider a $m$-variables semantic source $\left\{ X_1, X_2,...,X_m \right\}$ whose joint probabilistic distribution is modeled by BN. We assume that the order of $m$ variables is sorted according to their causal relations, i.e., a child node variable always follows its parent node variables.

\textbf{Theorem 1. (Lossless Compression of Semantic Sources)}  Give a $m$-variables source $\left\{ X_1, X_2,...,X_m \right\}$ with entropy  $H\left(  X_1, X_2,...,X_m  \right)$. For any code rate $R$ if $R > H\left( X_1, X_2,...,X_m \right)$, there exists a lossless source code for this source. 

\emph{Proof.} The proof of Theorem 1 easily follows from the proof of Shannon's first theorem \cite{b1}, which is omitted here.

\textbf{Remark 1.} By utilizing the conditional independence property of BN, the entropy of this source $H\left(  X_1, X_2,...,X_m  \right)$ can be written as
\begin{equation}
\begin{aligned}
   H({X_1},{X_2}, \ldots ,{X_m}) &=\sum\limits_{i = 1}^m {H\left( {{X_i}|{X_{i - 1}},...,{X_1}} \right)} \\&=\sum\limits_{i = 1}^m {H\left( {{X_i}|Parent({X_{i}})} \right)},
\end{aligned}
\label{entropy}
\end{equation}
where $Parent({X_{i}})$ denotes all parent variables of $X_{i}$. For an $m$-variables source, the rate of joint coding all the variables 
is always less than that of separate compression of each variable. Because 
\begin{equation}
\begin{aligned}
  &\sum\limits_{i = 1}^m {H\left( {{X_i}} \right)} - H({X_1},{X_2}, \ldots ,{X_m}) 
  \\&= \sum\limits_{i = 2}^m {I\left( {{X_i};Parent({X_{i}})} \right)}  \\ &\geqslant 0 .
\end{aligned}
\label{remark1.2}
\end{equation}
\textbf{Remark 2.}  If we directly compress samples generated by the source $\left\{ X_1, X_2,...,X_m \right\}$ with Huffman encoding, the time complexity is $\mathcal{O}\left( {k^mlogk^m} \right)$, where $k$ is the maximal number of states that each variable has. The computation overhead of sorting algorithm is infeasible when the $m$ is large. We can utilize the conditional relations between parents and child variables to significantly reduce the complexity of Huffman coding. specifically, we can iteratively sort and encode the samples starting from the root node. Then for each node, we utilize the conditional probability with regards to its parent nodes to sort and encode its samples. This avoids the time complexity increasing exponentially with the number of nodes. We assume the maximum number of parent variables among $\left\{ X_1, X_2,...,X_m \right\}$ is $L$ ($L$ is usually much smaller than $m$). Then the number of samples required to be coded each time is limited by $mk^{L}$. In this way, we have the overall complexity reduced from $\mathcal{O}\left( {{k^m}log{k^m}} \right)$
to $\mathcal{O}\left( mk^{L}logk^{L} \right)$.

\textbf{Theorem 2. (Lossy Compression of Semantic Sources)} 
The rate-distortion function for an $m$-variables semantic source whose joint distribution can be modeled by a BN, and distortion $D_1,D_2,...,D_m$ is given by
\begin{equation}
\begin{footnotesize}
\begin{aligned}
&{R_{{X_1},{X_2},...,{X_m}}}({D_1},{D_2},...,{D_m})\\&= \min _{\begin{subarray}{c}
  p({\hat x_1},{\hat x_2},...,{\hat x_m}|{x_1},{x_2},...,{x_m}) \\ 
  Ed({\hat x_1},{x_1}) \leqslant {D_1} \\ 
  Ed({\hat x_2},{x_2}) \leqslant {D_2} \\ 
  ...\\
  Ed({\hat x_m},{x_m}) \leqslant {D_m}
\end{subarray}} 
I({X_1},{X_2},...,{X_m};{\hat X_1},{\hat X_2},..,{\hat X_m}).
\end{aligned}
\label{rate-distortion}
\end{footnotesize}
\end{equation}
If  $R > R_{{X_1},{X_2},...,{X_m}}({D_1},{D_2},...,{D_m})$, there exists a lossy source code for this $m$-variables source at rate $R$ for a distortion not exceeding ${D_1},{D_2},...,{D_m}$. 

\emph{Proof.} This can be proved using a straightforward extension of Shannon's work \cite{b28}.

\textbf{Lemma 1.} For an $m$-variables semantic source whose joint distribution can be modeled by a BN, the rate-distortion function $R_{{X_1},{X_2},...,{X_m}}({D_1},{D_2},...,{D_m})$ can be bounded by 
\begin{equation}
\begin{footnotesize}
\begin{aligned}
\sum\limits_{i = 1}^m {{R_{{X_i}}}\left( {{D_i}} \right)} &\geqslant {R_{{X_1},{X_2},...,{X_m}}}({D_1},{D_2},...,{D_m}) \\&\geqslant \sum\limits_{i = 1}^m {{R_{{X_i}|Parent({X_{i}})}}\left( {{D_i}} \right)},
\end{aligned}
\label{rate-distortion lemma1}
\end{footnotesize}
\end{equation}
where ${{R_{{X_i}|Parent({X_{i}})}}\left( {{D_i}} \right)}$ represents the conditional rate-distortion function characterized by
\begin{equation}
\begin{footnotesize}
\begin{aligned}
&{{R_{{X_i}|Parent({X_{i}})}}\left( {{D_i}} \right)} 
\\&= \min _{\begin{subarray}{c}
  p({\hat x_i}|{x_i},Parent({x_{i}})) \\ 
  Ed({\hat x_i},{x_i})  \leqslant {D_i} \\ 
\end{subarray}} 
I({X_i};{\hat X_i}|Parent({X_{i}})),
\end{aligned}
\label{conditional rate-distortion}
\end{footnotesize}
\end{equation}
where
\begin{equation}
\begin{aligned}
   Ed({\hat x_i},{x_i}) = \sum\limits_{x_i,\hat x_i,Parent({x_{i}}),} \{p({\hat x_i}|{x_i},Parent({x_{i}}))\\p({x_i},Parent({x_{i}}))d({\hat x_i},{x_i})\}.
   \end{aligned}
\label{conditional rate-distortion1}
\end{equation}
\emph{Proof.} The upper bound in \eqref{rate-distortion lemma1} is a straightforward extension of the upper bound of Wyner and Ziv \cite{b29}, the proof of which is omitted here. The rigorous proof of the lower bound will be provided in a longer version.

\textbf{Remark 3.} The upper bound in Lemma 1 indicates that the rate of reconstructing all the variables within the given fidelity is always less than that of separate reconstruction of each variable. The rate-distortion of $m$-variables may be infeasible to obtain when $m$ is large. The lower bound in \eqref{rate-distortion lemma1} suggests that we can use the summation of conditional rate distortion to guide the design of lossy source coding instead.

\section{Lossy Compression of Correlated Semantic Elements With Side Information}
In semantic communications, the sender and receiver always have access to some background knowledge about the communication contents. This background knowledge can be used as side information to help the compression of intended messages. In this section, we study the compression of correlated semantic elements when side information exists at both the sender and receiver. We further evaluate the corresponding  rate-distortion function when the semantic elements follow  binary distribution and multi-dimensional Gaussian distribution respectively.

\textbf{Theorem 3. (Compression With Side Information)}
Given the bounded distortion measure $(d_1: \mathcal{X}_1 \times \hat{\mathcal{X}}_1 \to \mathcal{R}^+$,..., $d_m: \mathcal{X}_m \times \hat{\mathcal{X}}_m \to \mathcal{R}^+)$, where $\mathcal{R^+}$ denotes the set of nonnegative real numbers.
If some variable is observed and revealed to the encoder and decoder as side information, denoted by $Y$, the rate-distortion function for compressing the remaining variables $X_1, X_2,...,X_m$ is given by
\begin{equation}
\begin{aligned}
&{R_{{X_1},...,{X_m}|Y}}({D_1},...,{D_m})\\&= \min _{\begin{subarray}{c}
  p({\hat x_1},...,{\hat x_m}|{x_1},...,{x_m},{\text{y}}) \\ 
  Ed({\hat x_1},{x_1})  \leqslant {D_1} \\ 
  ...\\
  Ed({\hat x_m},{x_m}) \leqslant {D_m}
\end{subarray}} 
I({X_1},...,{X_m};{\hat X_1},..,{\hat X_m}|Y).
\end{aligned}
\label{conditional entropy}
\end{equation}
\emph{Proof}: We first prove the achievability of Theorem 3 by showing that for any rate $R  \ge {R_{{X_1},...,{X_m}|Y}}({D_1},...,{D_m})$, there exists a lossy source code with the rate $R$ and asymptotic distortion $(D_1,...,D_m)$. Let $p({\hat x_1},...,{\hat x_m}|{x_1},...,{x_m},{\text{y}})$ be the conditional probability that achieves equality in \eqref{conditional entropy} and satisfies the distortion requirements, i.e., $Ed({\hat x_1},{x_1}) \leqslant {D_1}$,…, $Ed({\hat x_m},{x_m}) \leqslant {D_m}$.  

 \emph{Generation of codebook}: Randomly generate a codebook $\mathcal{C}$ with the help of side information $Y$.  The codebook $\mathcal{C}$ consists of ${2^{nR}}$ sequence triples $(\hat x_1,...,\hat x_m)^n$ drawn i.i.d. according to $p({\hat x_1},...,{\hat x_m}|y)$, where $p({\hat x_1},...,{\hat x_m}|y) = \sum\nolimits_{{x_1},...,{x_m}} {p({x_1},...,{x_m}|y)p({{\hat x}_1},...,{{\hat x}_m}|{x_1},...,{x_m},y)} $. These codewords are indexed by $w \in \left\{ {1,2,...,{2^{nR}}} \right\}$. The codebook  $\mathcal{C}$ is revealed to both the encoder and decoder.

  \emph{Encoding and Decoding}: Encode the observing  $(x_1,...,x_m,y)^n$ by $w$ if its indexing sequence $(\hat x_1,...,\hat x_m)^n$ is distortion typical with $(x_1,...,x_m,y)^n$, i.e., $(x_1,...,x_m,y,\hat x_1,...,\hat x_m)^n \in T_\epsilon^n $. If there is more then one such index $w$, choose the least. If there is no such index, let $w=1$. After obtaining the index $w$, the receiver chooses the codeword $(\hat x_1,...,\hat x_m)^n$ indexed by $w$ to reproduce the sequence.

\emph{Calculation of distortion}: For an arbitrary codebook $\mathcal{C}$ and any $\epsilon > 0$, the sequences $(x_1,...,x_m)^n \in (X_1,...,X_m)^n$ can be divided into to two categories:

For one case. Sequences $(x_1,...,x_m,y)^n$ that is distortion typical with a codeword $(\hat x_1,...,\hat x_m)^n$  in the codebook $\mathcal{C}$, i.e., $d({\hat x_1},{x_1}) < {D_1}+\epsilon,..., d({\hat x_m},{x_m}) < {D_m}+\epsilon$. Because the total occurrence probability of such sequence is less than 1, the expected distortions contributed by these sequences are no more than $({D_1}+\epsilon,..., {D_m}+\epsilon)$.

For the second case. Sequences $(x_1,...,x_m,y)^n$ that there is no codeword in the codebook $\mathcal{C}$ that is distortion typical with $(x_1,...,x_m)^n$. The total occurrence probability of such sequences is denoted by $P_e$. Since the distortions for $(x_1,...,x_m)^n$ can be bounded by $(d_{max,1},...,d_{max,m})$, the expected distortions contributed by these sequences are no more than $(P_ed_{max,1},...,P_ed_{max,m})$, where the bounded distortion measure $d_{max}$ is defined by
\begin{equation}
d_{\max, i } \stackrel{\text { def }}{=} \max _{x_i \in \mathcal{X}_i, \hat{x}_i \in \hat{\mathcal{X}}_i} d(x_1, \hat{x}_i)<\infty .
\end{equation}

Hence the total distortions can be bound as
 \begin{equation}
 \begin{small}
\begin{aligned}
 Ed({\hat x_1},{x_1}) &\leqslant {D_1} + \epsilon+ P_ed_{max,1},\\ 
  &...\\
  Ed({\hat x_m},{x_m}) &\leqslant {D_m} + \epsilon+ P_ed_{max,m}.
\end{aligned}
\label{bounded distortions}
\end{small}
\end{equation}
If $P_e$ is small enough, the expected distortions are closed to $({D_1},..., {D_m})$.

\emph{The bound of $P_e$:}  The coding error probability can be bounded as
\begin{subequations} \label{error probability}
\begin{small}
\begin{align}
P_e & = \sum\limits_{(x_1,...,x_m,{y})^n} { {p\left( (x_1,...,x_m,y)^n \right) } }\nonumber \\& \cdot p\left\{ {((x_1,...,x_m,{y})^n,(\hat X_1,...,\hat X_m)^n_w) \notin T_\epsilon^n,\forall w \in \left[ {1:{2^{nR}}} \right]} \right\}  \hfill \label{22a}\\
   & \leqslant p\left\{ {\prod\limits_{w = 1}^{{2^{nR}}} {((x_1,...,x_m,{y})^n,(\hat X_1,...,\hat X_m)^n_w) \notin T_\epsilon^n} } \right\} \hfill \label{22b} \\
   & = {\left( {1 - p\left\{ {\left( {(x_1,...,x_m,{y},\hat X_1,...,\hat X_m)^n \in T_\epsilon^n} \right)} \right\}} \right)^{{2^{nR}}}} \hfill \label{22c} \\
   & \leqslant {\left( {1 - {2^{ - n\left[ {I({X_1},...,{X_n};{{\hat X}_1},...,{{\hat X}_n}|Y) + \delta (\epsilon)} \right]}}} \right)^{{2^{nR}}}} \hfill  \label{22d},
\end{align}
\end{small}
\end{subequations}
where  \eqref{22a} is obtained by applying the joint typicality theorem in \cite{b26}, \eqref{22b} follows from the fact that ${p\left( (x_1,...,x_m,y)^n \right) }$ is at most 1, and \eqref{22c} and \eqref{22d} are obtained though the property of joint typical sequence. We note that  $\left( 1-z \right)^t\leqslant e^{\left( -tz \right)}$ for $z\in\left[ 0,1 \right]$ and $0\leqslant{t}$, and \eqref{error probability} can be rewritten as 
\begin{equation} 
\begin{aligned}
P_e  \leqslant \exp \left( {{2^{ - n\left[ {R - I({X_1},...,{X_n};{{\hat X}_1},...,{{\hat X}_n}|Y) - \delta (\epsilon)} \right]}}} \right) \hfill ,
\end{aligned}
\label{error probability2}
\end{equation}
where $\delta (\epsilon)\ \to 0$ when $n \to \infty $. We note that $P_e$ goes to zero with $n$ if ${R >  I({X_1},...,{X_n};{{\hat X}_1},...,{{\hat X}_n}|Y) + \delta (\epsilon)}$. This proves the rate-distortion pairs $\left(R, D_1,...,D_m \right)$ is achievable if ${R > R\left(D_1,...,D_m \right)}$.

We then prove the converse of Theorem 3 by showing that for any source code meeting the distortion requirements $(D_1,...,D_m)$, then the rate $R$ of the code must satisfy  $R  \ge {R_{{X_1},...,{X_m}|Y}}({D_1},...,{D_m})$. We consider any $\left(n, 2^{nR}\right)$ code with an encoding function $f_n: ({\mathcal{X}_1},...,{\mathcal{X}_m}, {\mathcal{Y}})^n \to \left\{ 1,2,...,2^{nR} \right\} $. Then we have 
\begin{subequations} \label{converse1}
\begin{footnotesize}
\begin{align}
  nR &\geqslant H\left( {{f_n}\left( ({X_1,...,X_m,{Y}})^n \right)} \right) \label{24a}\\&\geqslant H\left( {{f_n}\left( ({X_1,...,X_m,{Y}})^n \right)|{Y^n}} \right) \hfill \label{24b} \\
   &\geqslant H\left( {{f_n}\left( ({X_1,...,X_m,{Y}})^n \right)|{Y^n}} \right) \nonumber \\
   &- H\left( {{f_n}\left( ({X_1,...,X_m,{Y}})^n \right)|({X_1,...,X_m,{Y}})^n} \right) \hfill \label{24c} \\
   &\geqslant I\left( {({X_1,...,X_m})^n;(\hat X_1,...,\hat X_m)^n|{Y^n}} \right) \hfill \label{24d} \\
   &= I\left( {({X_1,...,X_m,{Y}})^n;(\hat X_1,...,\hat X_m)^n} \right) \nonumber \\&- I\left( {{Y^n};(\hat X_1,...,\hat X_m)^n} \right) \hfill  \label{24e} ,
\end{align}
\end{footnotesize}
\end{subequations}
where \eqref{24a} follows from the fact that the number of codewords is $2^{nR}$, \eqref{24b} is obtained by the fact that conditioning reduces entropy, \eqref{24c} is obtained by introducing a nonnegative term, \eqref{24d} follows from the property of data-processing, and \eqref{24e} follows from the property of conditional mutual information. By applying the chain rule of mutual information to \eqref{24e}, we have
\begin{subequations} \label{converse2}
\begin{footnotesize}
\begin{align}
  &nR \nonumber \\&\geqslant \sum\limits_{i = 1}^n I\left( {{X_{1,i}},...,{X_{m,i}},{Y_i};(\hat X_1,...,\hat X_m)^n|X_{1}^{i - 1},} {...,X_{m}^{i - 1},Y^{i - 1}} \right)\nonumber  \\& - \sum\limits_{i = 1}^n {I\left( {{Y_i};(\hat X_1,...,\hat X_m)^n|Y^{i - 1}} \right)} \label{25a} \\&=
   \sum\limits_{i = 1}^n {H\left( {{X_{1,i}},...,{X_{m,i}},{Y_i}|X_{1}^{i - 1},...,X_{m}^{i - 1},Y^{i - 1}} \right)}  \nonumber \\&- \sum\limits_{i = 1}^n {H\left( {{X_{1,i}},...,{X_{m,i}},{Y_i}|(\hat X_1,...,\hat X_m)^n,X_{1}^{i - 1},...,X_{m}^{i - 1},Y^{i - 1}} \right)}  \nonumber \\&- \sum\limits_{i = 1}^n {H\left( {{Y_i}|Y^{i - 1}} \right)}  + \sum\limits_{i = 1}^n {H\left( {{Y_i}|(\hat X_1,...,\hat X_m)^n,Y^{i - 1}} \right)} \label{25b} \hfill \\
   &= \sum\limits_{i = 1}^n {H\left( {{X_{1,i}},...,{X_{m,i}},{Y_i}} \right)}  - \sum\limits_{i = 1}^n {H\left( {{Y_i}} \right)} \nonumber \nonumber\\& - \sum\limits_{i = 1}^n {H\left( {{Y_i}|(\hat X_1,...,\hat X_m)^n} \right)} + \sum\limits_{i = 1}^n {H\left( {{Y_i}|(\hat X_1,...,\hat X_m)^n} \right)}  \nonumber\\&- \sum\limits_{i = 1}^n {H\left( {{X_{1,i}},...,{X_{m,i}}|{Y_i},(\hat X_1,...,\hat X_m)^n} \right)}  \label{25d}\hfill \\
  & \geqslant \sum\limits_{i = 1}^n {H\left( {{X_{1,i}},...,{X_{m,i}}|{Y_i}} \right)}  \nonumber \\&- \sum\limits_{i = 1}^n {H\left( {{X_{1,i}},...,{X_{m,i}}|\hat X_{1,i}^{},...,\hat X_{m,i}^{},{Y_i}} \right)} \label{25d1} \hfill 
\\&= \sum\limits_{i = 1}^n {I\left( {{X_{1,i}},...,{X_{m,i}};\hat X_{1,i}^{},...,\hat X_{m,i}^{}|{Y_i}} \right)} \nonumber \hfill
\\& \geqslant \sum\limits_{i = 1}^n {R\left( {E{d_1}\left( {{X_{1,i}},\hat X_{1,i}^{}} \right),...,E{d_m}\left( {{X_{m,i}},\hat X_{m,i}^{}} \right)} \right)}  \label{25e} \hfill
\\& \geqslant nR\left( {E{d_1}\left( {X_1^n,\hat X_1^n} \right),...,E{d_m}\left( {{X_{m,i}^n},\hat X_m^n} \right)} \right) \nonumber \hfill 
\\& \geqslant nR\left( {{D_1},...,{D_m}} \right)\nonumber,
\end{align}
\end{footnotesize}
\end{subequations}
where $X_{j}^{i - 1}$ denotes the sequence $(X_{j,1},...,X_{j,i-1})$, \eqref{25b} follows from the definition of conditional mutual information. \eqref{25d} follows from the chain rule and the fact that the source is memoryless, i.e., $({X_{1,i}},...,{X_{m,i}},{Y_i})$ and  $(X_{1}^{i - 1},...,X_{m}^{i - 1},Y^{i - 1})$ are independent. \eqref{25d1} is obtained by the fact that conditioning reduces entropy. And \eqref{25e} follows from the definition of $R\left({D_1},...,{D_m}\right)$. This proves the converse of Theorem 3.

\textbf{Lemma 2.} 
Given the known variables $Y$, if an $m$-variables source can be divided into several conditional independent subsets $\mathcal{V}_1,...\mathcal{V}_l$ by the property of BN, then
\begin{equation} 
\begin{footnotesize}
\begin{aligned}
R_{{{\mathcal{V}_1},...,{\mathcal{V}_l}|Y}}({D_{1}},...,{D_{m}}) = \sum\limits_{i = 1}^l {{R_{{\mathcal{V}_i}|Y}}\left( {{D_{j}}}, j\in{\mathcal{V}_i}\right)} ,
\end{aligned}
\label{lemma2}
\end{footnotesize}
\end{equation}
where the term $R_{\mathcal{V}_i|Y}\left(D_{{j, j\in{\mathcal{V}_i}}}\right)$ is given by:
\begin{equation} 
\begin{footnotesize}
\begin{aligned}
{R_{{{\mathcal{V}}_i}|Y}}({D_{{j, j\in{\mathcal{V}_i}}}}) = \min _{\begin{subarray}{c}
  p({\hat v_i}|{v_i},{{y}}):\\Ed_{j}({\hat x_j},{x_j}) \leqslant {D_{j},j\in \mathcal{V}_i}
\end{subarray}} 
I({\mathcal{V}_i};{\hat {\mathcal{V}}_i}|Y).
\end{aligned}
\label{lemma2_1}
\end{footnotesize}
\end{equation}

\emph{Proof}: The proof of Lemma 2 will be provided in a longer version.

\textbf{Remark 4.} Lemma 2 implies that if a set of semantic elements  can be divided into several conditional independent subsets by using the property of BN with side information $Y$, compressing the source variable set jointly is the same as compressing these conditional independent subsets separately in terms of the distortions and rate. We note that the separate compression of conditional independent subsets can significantly reduce the complexity of coding.

\textbf{Example 1.}
\begin{figure}{}
\centering
\subfigure[]{\includegraphics[width=2.5cm]{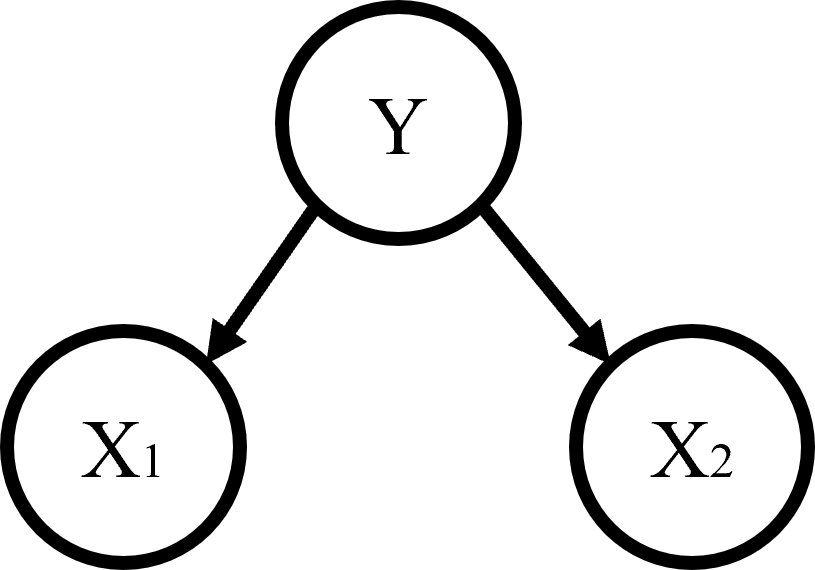}} 
\subfigure[]{\includegraphics[width=2.5cm]{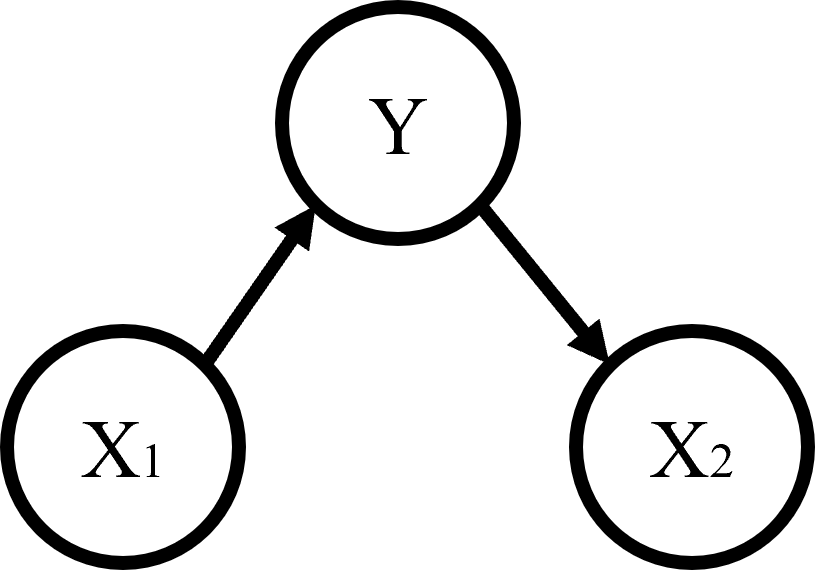}}
\caption{Two examples on semantic sources.} 
\label{fig111}
\end{figure}
Consider two different sources as shown in Fig.~\ref{fig111}.  The characteristics of BN indicate that the variables  $X_1$ and $X_2$ in both cases of Fig.~\ref{fig111} are conditional independent given $Y$. By Lemma 2, we have that if the variable $Y$ is revealed to the encoder and decoder as side information, then
\begin{equation} 
\begin{footnotesize}
\begin{aligned}
R\left( D_1, D_2 \right) &= \min _{\begin{subarray}{c}
  p({\hat x_1},{\hat x_2}|{x_1},{x_2},{\text{y}}) \\ 
  Ed({\hat x_1},{x_1}) \leqslant {D_1} \\ 
  Ed({\hat x_2},{x_2}) \leqslant {D_2}
\end{subarray}} 
I({X_1},{X_2};{\hat X_1},{\hat X_2}|Y) \\&= R_{X_1|Y}\left(D_1\right)+R_{X_2|Y}\left(D_2\right).
\end{aligned}
\label{example 4}
\end{footnotesize}
\end{equation}

\textbf{Example 2.} We first explore the rate-distortion function of a binary semantic source given side information with Hamming distortion measure. This semantic source consists of three semantic elements $(X_1, X_2, Y)$ whose probabilistic distribution can be modeled by a BN as shown in Fig.~\ref{fig111}(a). The inter-variable dependence structures  $(X_1,Y)$ and $(X_2,Y)$ are doubly symmetric binary distributed with parameters $p_1$ and $p_2$ respectively, where 
\begin{equation}
\small{
\begin{aligned}
p\left( {{x_1},y} \right) = \left[ {\begin{array}{*{20}{c}}
  {\frac{{1 - {p_1}}}{2}}&{\frac{{{p_1}}}{2}} \\ 
  {\frac{{{p_1}}}{2}}&{\frac{{1 - {p_1}}}{2}} 
\end{array}} \right], p\left( {{x_2},y} \right) = \left[ {\begin{array}{*{20}{c}}
  {\frac{{1 - {p_2}}}{2}}&{\frac{{{p_2}}}{2}} \\ 
  {\frac{{{p_2}}}{2}}&{\frac{{1 - {p_2}}}{2}} 
\end{array}} \right].
\end{aligned}}
\label{example binary1}
\end{equation}
 By summing the joint probability distribution over all values of $x_1$ and $x_2$, we can obtain the marginal distribution $p(y)$. The the conditional distributions  $p(x_1|y)$ and $p(x_2|y)$ can be obtained through Bayesian criterion as
\begin{equation}
\begin{footnotesize}
\begin{aligned}
p\left( {{x_1}|y} \right) = \left[ {\begin{array}{*{20}{c}}
  {{1 - {p_1}}}&{{{p_1}}} \\ 
  {{{p_1}}}&{{1 - {p_1}}} 
\end{array}} \right], p\left( {{x_2}|y} \right) = \left[ {\begin{array}{*{20}{c}}
  {{1 - {p_2}}}&{{{p_2}}} \\ 
  {{{p_2}}}&{{{1 - {p_2}}}} 
\end{array}} \right].
\end{aligned}
\label{example binary2}
\end{footnotesize}
\end{equation}
By Lemma 2, we have $R\left( D_1, D_2 \right) = R_{X_1|Y}\left(D_1\right) + R_{X_2|Y} \left(D_2\right)$. Following the conditional rate-distortion function of binary sources in \cite{b27}, it yields 
\begin{subequations}\label{example binary3}
\begin{footnotesize}
\begin{align}
R_{X_1|Y}\left(D_1\right)=\left[h_b(p_1)-h_b(D_1) \right]_{0 \leqslant D_1 \leqslant p_1 },\label{35a}\\
R_{X_2|Y}\left(D_2\right)=\left[h_b(p_2)-h_b(D_2) \right]_{0 \leqslant D_2 \leqslant p_2 }\label{35b}.
\end{align}
\end{footnotesize}
\end{subequations}
Thus,
\begin{equation}
\begin{footnotesize}
\begin{aligned}
R\left( D_1, D_2 \right)&=\left[h_b(p_1)-h_b(D_1) \right]_{0 \leqslant D_1 \leqslant p_1 }\\&+\left[h_b(p_2)-h_b(D_2) \right]_{0 \leqslant D_2 \leqslant p_2 }.
\end{aligned}
\label{example binary3}
\end{footnotesize}
\end{equation}

We then consider the conditional rate-distortion function of a  Gaussian source whose probabilistic distribution can be modeled by a BN as shown in Fig.~\ref{fig111}(a). We use the mean-squared-error distortion measure here. $p\left( {{x_1},y} \right)$ is two-dimensional Gaussian distribution with parameters $m_{X_1}$, $m_Y$, $\sigma_{X_1}$, $\sigma_{Y},r_1$ as
\begin{equation}
\begin{footnotesize}
\begin{aligned}
p\left( {{x_1},y} \right) & =\frac{1}{2 \pi \sigma_{X_1} \sigma_Y \sqrt{1-{r_1}^2}} \exp \left\{-\frac{1}{2 \sigma_{X_1}^2 \sigma_Y^2\left(1-{r_1}^2\right)}\right\} \\
& \cdot\left\{\left(\frac{x_1-m_{X_1}}{\sigma_{X_1}}\right)^2+\left(\frac{y-m_Y}{\sigma_Y}\right)^2\right. \\
& \left.\quad-2 r \frac{\left(x_1-m_{X_1}\right)\left(y-m_Y\right)}{\sigma_{X_1} \sigma_Y}\right\}. 
\end{aligned}
\label{example Gaussian1}
\end{footnotesize}
\end{equation}
The conditional distribution $p\left( {{x_1}|y} \right)$ is also Gaussian distribution as
\begin{equation}
\begin{footnotesize}
\begin{aligned}
p\left( {{x_1}|y} \right)&=\left(2 \pi \sigma_{X_1}^2\left(1-{r_1}^2\right)\right)^{-1 / 2} \exp \left\{-\left(2 \sigma_{X_1}^2\left(1-r^2\right)\right)^{-1}\right. \\
& \left.\cdot\left[x_1-m_{X_1}-r_1 \frac{\sigma_{X_1}}{\sigma_Y}\left(y-m_Y\right)\right]^2\right\}.
\end{aligned}
\label{example Gaussian2}
\end{footnotesize}
\end{equation}
Therefore, we can obtain the rate-distortion function $R_{X_1|Y}(D_1)$ according to Shannon's work \cite{b28}
\begin{equation}
\begin{footnotesize}
\begin{aligned}
R_{X_1|Y}(D_1)=\left[\frac{1}{2} \log \frac{\sigma_{X_1}^2\left(1-{r_1}^2\right)}{D_1}\right]_{ 0 \leq D_1 \leq \sigma_{X_1}^2\left(1-{r_1}^2\right)}.
\end{aligned}
\label{example Gaussian3} 
\end{footnotesize}
\end{equation}
Similarly, we assume $p\left( {{x_2},y} \right)$ also follows two-dimensional Gaussian distribution with parameters $m_{X_2}$, $m_Y$, $\sigma_{X_2}$, $\sigma_{Y},r_2$, and  $R_{X_2|Y}(D_2)$ is given by
\begin{equation}
\begin{footnotesize}
\begin{aligned}
R_{X_2|Y}(D_2)=\left[\frac{1}{2} \log \frac{\sigma_{X_2}^2\left(1-{r_2}^2\right)}{D_1}\right]_{ 0 \leq D_2 \leq \sigma_{X_2}^2\left(1-{r_2}^2\right)}.
\end{aligned}
\label{example Gaussian4}
\end{footnotesize}
\end{equation}
By Lemma 2, we can obtain $R\left( D_1, D_2 \right)$
\begin{equation}
\begin{footnotesize}
\begin{aligned}
R\left( D_1, D_2 \right)&=\left[\frac{1}{2} \log \frac{\sigma_{X_1}^2\left(1-{r_1}^2\right)}{D_1}\right]_{ 0 \leq D_1 \leq \sigma_{X_1}^2\left(1-{r_1}^2\right)} \\&+\left[\frac{1}{2} \log \frac{\sigma_{X_2}^2\left(1-{r_2}^2\right)}{D_1}\right]_{ 0 \leq D_2 \leq \sigma_{X_2}^2\left(1-{r_2}^2\right)}.
\end{aligned}
\label{example Gaussian2}
\end{footnotesize}
\end{equation}

\section{Conclusion }

In this paper, we investigated compression of a semantic source which consists a set of correlated semantic elements, the joint probabilistic distribution of which can be modeled by a BN. Then we derived the theoretical limits on lossless compression and lossy compression of this semantic source, as well as the lower and upper bounds on the rate-distortion function. We also investigated the lossy compression problem of the semantic source with side information at both the encoder and decoder. We further proved that the conditional rate distribution function is equivalent to the summation of conditional rate distribution function of each conditionally independent set of variables given the side information. We also derived the conditional rate-distortion functions when the semantic elements of source are binary distribution and multi-dimensional distribution, respectively.

\end{document}